\documentclass[conference]{IEEEtran}
\IEEEoverridecommandlockouts
\usepackage{cite}
\usepackage{amsmath,amssymb,amsfonts}
\usepackage{algorithmic}
\usepackage{graphicx}
\usepackage{textcomp}
\usepackage{xcolor}
\usepackage{etoolbox}

\usepackage{textcomp}
\usepackage{xcolor}
\usepackage{float}

\usepackage{xcolor,soul}
\usepackage{xspace}
\usepackage{enumitem}
\usepackage{titlecaps}
\usepackage{tabularx}
\usepackage{balance}
\usepackage{color}
\usepackage{comment}
\usepackage{graphicx}
\usepackage{grffile} 
\usepackage{subcaption}

\usepackage{url}
\usepackage{hyperref}

\usepackage[left=0.62in,right=0.75in,top=0.75in, bottom=1in, columnsep=0.21in]{geometry}

\usepackage{setspace}
\def\BibTeX{{\rm B\kern-.05em{\sc i\kern-.025em b}\kern-.08em
    T\kern-.1667em\lower.7ex\hbox{E}\kern-.125emX}}

\newcommand{\sol}{{\em DarkHorse}\xspace}

\newcommand{\eg}{{\em e.g.,}\ }
\newcommand{\ie}{{\em i.e.,}\ }

\begin{document}

\makeatletter
\def\ps@IEEEtitlepagestyle{%
  \def\@oddfoot{\mycopyrightnotice}%
  \def\@evenfoot{}%
}
\def\mycopyrightnotice{%
  {\footnotesize \textcolor{red}{\begin{tabular}[t]{@{}l@{}} This paper has been accepted for publication by the 48th IEEE Conference on Local Computer Networks (LCN). © 2023 IEEE. Personal use \\ of this material is permitted. Permission from IEEE must be obtained for all other uses, in any current or future media, including reprinting/republishing \\ this material for advertising or promotional purposes, creating new collective works, for resale or redistribution to servers or lists, or reuse of any copyrighted \\ component of this work in other works.\end{tabular}}}
  \gdef\mycopyrightnotice{}
}

\title{\sol: A UDP-based Framework to Improve the Latency of Tor Onion Services}

\author{
\IEEEauthorblockN{Md Washik Al Azad}
\IEEEauthorblockA{University of Notre Dame\\
malazad@nd.edu}
\and
\IEEEauthorblockN{Hasniuj Zahan}
\IEEEauthorblockA{University of Nebraska at Omaha\\
hzahan@unomaha.edu}
\and
\IEEEauthorblockN{Sifat Ut Taki}
\IEEEauthorblockA{University of Notre Dame\\
staki@nd.edu}
\and
\IEEEauthorblockN{Spyridon Mastorakis}
\IEEEauthorblockA{University of Notre Dame \\
mastorakis@nd.edu}
}



\setlength{\skip\footins}{4pt}

\maketitle

\begin{abstract}
Tor is the most popular anonymous communication overlay network which hides clients’ identities from servers by passing packets through multiple relays. To provide anonymity to both clients and servers, Tor onion services were introduced by increasing the number of relays between a client and a server. Because of the limited bandwidth of Tor relays, large numbers of users, and multiple layers of encryption at relays, onion services suffer from high end-to-end latency and low data transfer rates, which degrade user experiences, making onion services unsuitable for latency-sensitive applications. In this paper, we present a UDP-based framework, called \sol, that improves the end-to-end latency and the data transfer overhead of Tor onion services by exploiting the connectionless nature of UDP. Our evaluation results demonstrate that \sol is up to 3.62$\times$ faster than regular TCP-based Tor onion services and reduces the Tor network overhead by up to 47\%.

\end{abstract}

\begin{IEEEkeywords}
Tor, Onion Services, Anonymous Communication, Latency, UDP
\end{IEEEkeywords}

\section {Introduction}
\label{sec:intro}
Maintaining privacy while accessing the Internet has been a major concern over the past several years.
Researchers have tried to address this problem by proposing various network anonymization techniques, such as Mix-Net~\cite{mixnet}, Babel~\cite{babel}, and Mixminion~\cite{mixminion}. However, these proposals were not widely adopted because of the issues with impractically high latency. The Onion Routing project (Tor)~\cite{dingledine2004tor} was able to provide anonymous services to anonymous users while achieving substantially lower latency than the previous approaches. As such, Tor established itself as the most popular low-latency anonymous network service to this day.

Global Internet traffic is growing rapidly year over year. According to the CISCO Annual Internet Traffic Report, the fixed broadband Internet speed and the mobile cellular Internet speed will reach 110.4 Mbps and 43.9 Mbps, respectively, by the end of 2023~\cite{forecast2020cisco}. The massive increase in Internet speed allows service providers to deploy applications over the Internet that were not previously possible, such as video streaming, Augmented Reality/Virtual Reality (AR/VR), robotics, and healthcare applications~\cite{sandvine2023global}. The requirement for 
low latency networks is expected to only grow in the future. 


Although Tor offers a relatively low-latency anonymous overlay network, the overall performance and data transfer rates are still low. Tor works by deploying a number of relay nodes around the world and routing data through these nodes via multiple layers of encryption to achieve anonymity. Routing data through the relay nodes incurs an overhead on the network, which increases with the number of relay nodes. Moreover, the relay nodes have limited computation power and bandwidth, which further impacts the performance of Tor.

In 2003, Tor onion services (also known as hidden services) were introduced \cite{tor_intro} to provide anonymity to both clients and servers at the same time by doubling the number of relays between two communicating parties \cite{tor_spec}. As a result, the latency of onion services increased further, making onion services impractical to use for 
low-latency network applications (\eg video streaming over Tor). Furthermore, transferring large data files through onion services is unreliable due to low data transfer rates, availability, and unpredictable performance of Tor relays. As such, reducing the number of relay nodes between the client and the onion server can improve the overall performance and the limited resources of the overall Tor overlay network can be effectively utilized. And, as a result, the capacity of the Tor network to serve the total number of clients will increase.


To address these limitations, in this paper, we propose \sol, a UDP-based framework for onion services. 
\sol exploits the connectionless nature of UDP to create a unidirectional path from a sender to a receiver through the Tor overlay network using a temporary source IP address for transmitting packets. As a result, \sol enables onion services to use 50\% fewer relay nodes, improving the performance and latency of the network while preserving the anonymity of both clients and servers. The contributions of our paper are the following:

\begin{itemize}
    \item We present the design of \sol, which improves the performance and latency of onion services.
    \item We develop a \sol prototype and evaluate its performance by comparing (1) bootstrap time, (2) end-to-end per packet delay, (3) data transfer time, and (4) overhead with the vanilla onion service design\footnote{In this paper, we use the term ``vanilla Tor'' to refer to regular TCP-based Tor and the term ``vanilla onion services'' to refer to regular TCP-based Tor onion services.}.
    \item We scale our evaluation up to five hundred concurrent client connections to analyze how the performance of \sol is impacted as the number of concurrent connections grows.
\end{itemize}

The rest of the paper is organized as follows: Section~\ref{sec:back} discusses previous work that aimed to improve the performance of the Tor network. Section~\ref{sec:design} presents the design and workflow of \sol. Section~\ref{sec:eval} presents our evaluation results and comparison with vanilla onion services. Finally, Section~\ref{sec:conclusion} concludes our paper and discusses future work.   
\section{Related Work}
\label{sec:back}

Tor \cite{dingledine2004tor} is a realization of onion routing \cite{reed1998anonymous} to provide anonymity to clients on the public Internet through an overlay network. Tor uses multiple relay nodes (called a circuit, usually, with three nodes: a guard node, a middle node, and an exit node) to send packets from clients to servers in order to hide the identity of clients. To ensure the anonymity of both clients and servers, onion services (formerly known as hidden services) \cite{dingledine2004tor} were later introduced, where additional relay nodes are used to offer server anonymity. As the relay nodes are spread all over the world with limited bandwidth and multiple layers of encryption, the anonymity provided by onion services comes at the cost of high latency \cite{dingledine2009performance,alsabah2016performance}. 

Different approaches have been proposed to improve the performance of the Tor network from various perspectives, such as efficient path (relay) selection, multi-path routing, and multi-threaded relays. Sherr \textit{et al.} \cite{sherr2009scalable} and Panchenko \textit{et al.} \cite{panchenko2009path} proposed two path selection algorithms based on the measured latency between two endpoints. In ShorTor \cite{hogan2022shortor}, the authors exploited multi-hop overlay routing of Content Delivery Networks (CDNs) to find an optimal path between users and servers with the goal of reducing the latency. Multi-path routing-based approaches were proposed in \cite{yang2015mtor} and \cite{de2020out} to improve the performance of Tor for bandwidth-intensive applications. A multi-threaded internal architecture for relays has also been proposed, so that resource utilization is enhanced and, as a result, the throughput of the relays and the capacity of the network are increased~\cite{engler2021weaving}.

Finally, a UDP-based approach, called UDP-OR, has been proposed \cite{viecco2008udp}, where a client is connected to an onion proxy using TCP, connections between two intermediate relays of a circuit use UDP, and the exit node communicates with a server using TCP. Although UDP-OR can be used for onion services, it requires at least six relay nodes (for a standard onion service connection) between clients and servers in order to provide anonymity to both communicating parties. To this end, our proposed framework, \sol, can preserve the anonymity of both clients and servers by using only three relay nodes for onion services.
\section{\sol Design}
\label{sec:design}

\subsection{Design Assumptions and Overview}

We assume that Tor relays, clients, and onion servers allow UDP traffic apart from regular TCP-based Tor traffic. 
We assume that a client can be anonymized from a server with a circuit consisting of $n$ Tor relays, and with another set of $m$ Tor relays, the server can also be anonymized from the client. As a result, to provide anonymity to both communicating parties (\ie for an onion service), $(n + m)$ Tor relays are needed in total. In  general, $m$ is equal to $n$ to provide the same level of anonymity to both clients and servers, and $n$ and $m$ are set to three, since this is considered as a good balance between performance and anonymity by the Tor community~\cite{tor_spec}. All attacks, which are possible against vanilla Tor and vanilla onion services (such as traffic analysis and correlations, timing attack, and fingerprinting attack), are also possible against \sol. For our proposed framework, there is no need to modify the existing relay selection algorithms for Tor circuits. We do not claim to improve the anonymity of the client and the server with \sol as compared to the vanilla onion service. Instead, our proposed framework improves the performance of onion services in terms of data transfer rates and latency by reducing the number of required relays by half compared to vanilla onion services. 
As a result, the resource utilization of the Tor network, such as bandwidth and computing power of relays, for data transfers will be reduced. Finally, \sol is only applicable to onion services (where the anonymity of both clients and servers is required) instead of regular Tor with three relays (where only the client's anonymity is required).

\begin{figure}[]
 \centering
 \includegraphics[width=1\columnwidth]{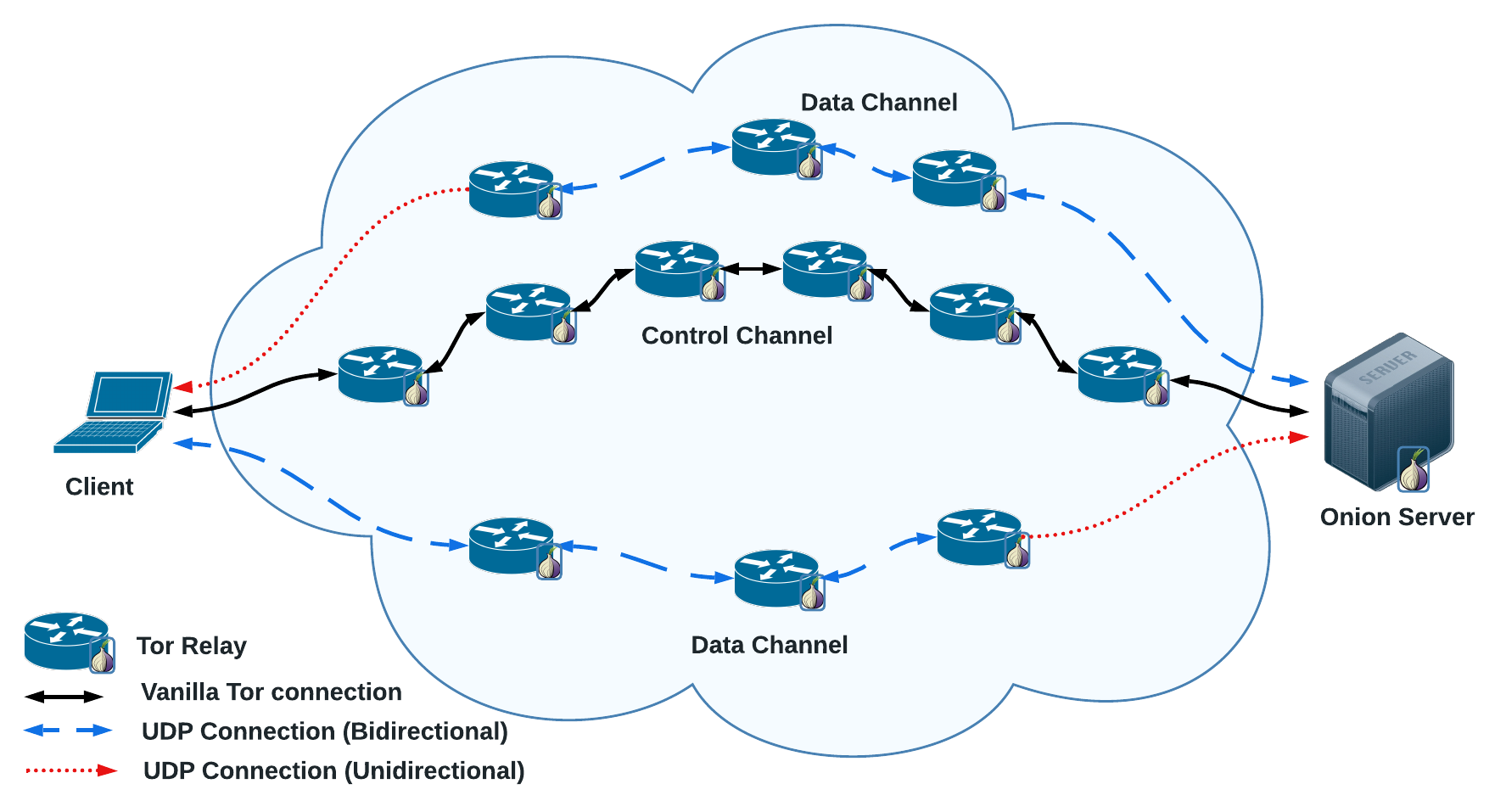}
 \caption{An overview of connections among Tor relays in \sol.} 
 \label{Figure:Design_overview}
\vspace{-0.5cm}
\end{figure}

\begin{figure}[]
 \centering
 \includegraphics[width=1\columnwidth]{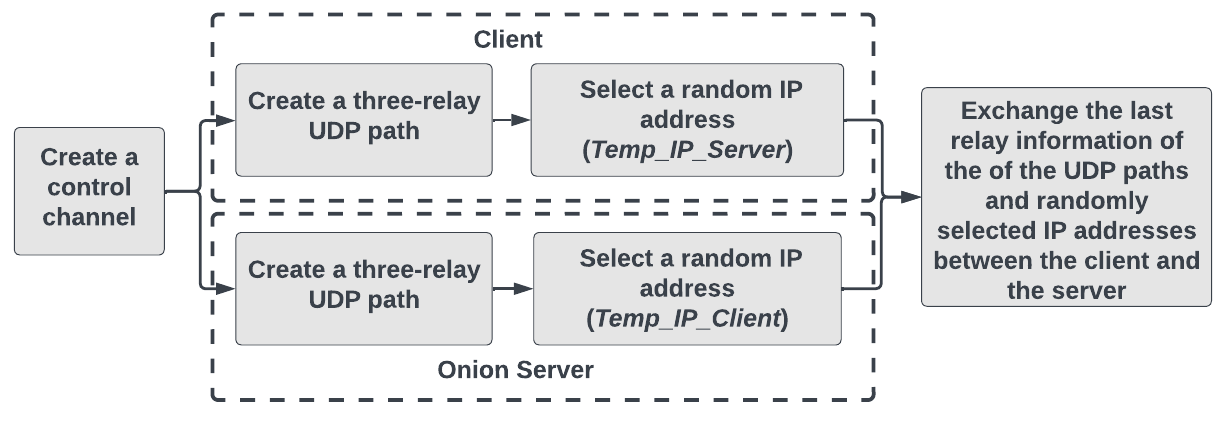}
 \caption{The workflow of \sol's bootstrap phase.} 
 \label{Figure:bootstrap}
 \vspace{-0.5cm}
\end{figure}

Figure \ref{Figure:Design_overview} shows the connections between a client and a server in \sol. These connections are divided into two types of channels: (i) control channels; and (ii) data channels. A control channel is a session between the client and the onion server through vanilla onion service and will be used to exchange information that is needed in order to create data channels. A data channel is a one-way UDP connection (\ie packets flow only in one direction) either from the client to the server or vice versa. To make a UDP connection unidirectional, the sender of the packets will use a temporary IP address (instead of its actual IP address) as the source IP address. As a result, the receiver of the packets cannot reach the sender using the same path. The reason we do not use TCP for unidirectional connections is that TCP requires a 3-way handshake between the sender and the receiver. There are two separate data channels in \sol: one channel for sending packets from the client to the onion server and another one for sending packets in the opposite direction (\ie onion server to client).

The operation of \sol has two phases: (i) bootstrap phase; and (ii) data transfer phase. In the bootstrap phase, the control channel and the data channels are created. In the data transfer phase, data packets are transmitted and lost packets are recovered (retransmitted).

\subsection{Bootstrap Phase of \sol}

Figure \ref{Figure:bootstrap} shows the workflow of the bootstrap phase. It starts with creating a control channel between the client and the server. After creating the control channel, both the client and the server select three relay nodes each to establish data channels. Tor's relay selection algorithm is used to choose these relays of the data channels. Similar to the TCP-based Tor circuit creation process, encryption keys are exchanged with these selected relays to build a UDP-based circuit and perform onion routing through that circuit. Then two random IP addresses \textit{Temp\_IP\_Server} and \textit{Temp\_IP\_Client} are selected by the client and the server respectively. Once these temporary IP addresses are selected, the client and the server exchange the IP address of the last relay of each data channel and the temporary IP addresses, so that data channels can be created. We discuss mechanisms to achieve the selection of temporary IP addresses in Section~\ref{sec:discussion}.

\begin{figure}[]
 \centering
 \includegraphics[width=1\columnwidth]{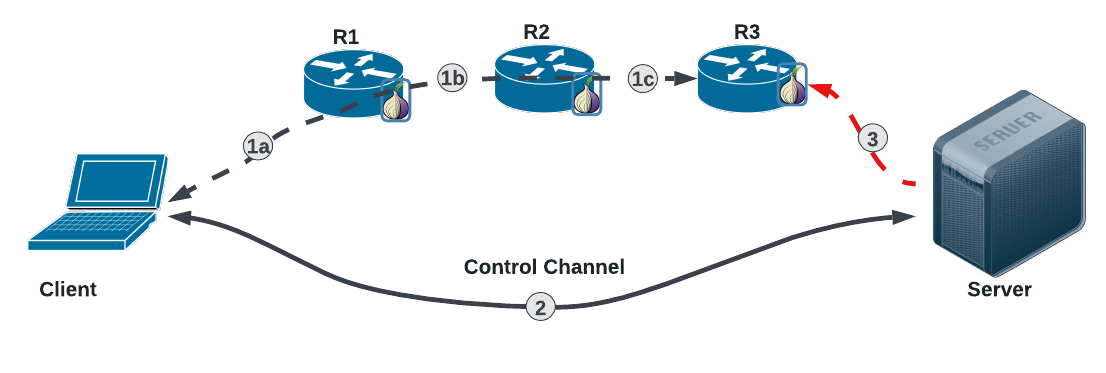}
 \caption{An example of creating data channel from a server to a client.} 
 \label{Figure:data_channel}
 \vspace{-0.5cm}
\end{figure}

Figure \ref{Figure:data_channel} shows the steps to create a data channel from a server to a client. First, the client selects three Tor relays (R1, R2, and R3) and creates a path. The client also selects a random IP address (\textit{Temp\_IP\_Server}). The server will use \textit{Temp\_IP\_Server} as a temporary IP address to send UDP packets toward the client through the path consisting of R1, R2, and R3. Since \textit{Temp\_IP\_Server} is not the actual IP address of the server, the client cannot reach back to the server using the same path. After selecting the IP addresses, the client sends the IP address of R3 and \textit{Temp\_IP\_Server} to the server. Finally, after receiving these IP addresses, the server begins to send UDP datagrams to R3 using \textit{Temp\_IP\_Server} as the source IP address.

\subsection{Data Transfer Phase of \sol}
\begin{figure}[]
 \centering
 \includegraphics[width=0.9\columnwidth]{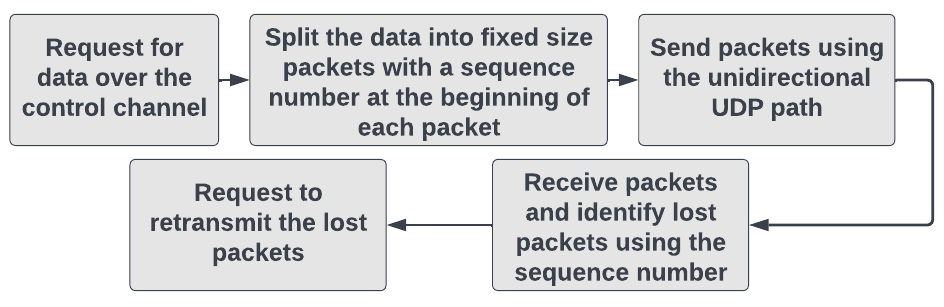}
 \caption{The workflow of \sol's data transfer phase.} 
 \label{Figure:data_transmit}
\end{figure}

Figure \ref{Figure:data_transmit} shows the operation workflow of transferring data packets using a data channel in \sol. It consists of five steps as follows.

\noindent\textbf{Step 1:} The first step of the data transfer process is to send a request for data. \sol supports two modes of sending requests. A request can be sent either over the control channel or over the data channel. If a request is sent over the control channel, then it can be an HTTP request. Otherwise, it is a UDP-based variant of HTTP request (such as an HTTP/3 request \cite{http3}) when sent over the data channel. Each mode of sending requests has its own advantages and disadvantages. Sending a request over the data channel is faster compared to sending a request over the control channel. However, due to the unreliable nature of UDP, sometimes a request cannot reach its destination. In this case, re-sending a request will be needed, which will lead to additional delay. On the other hand, using the control channel to send requests can be slow as the requests have to go through multiple Tor relays around the world. Nevertheless, unlike a data channel, packet losses and retransmissions will be handled by TCP, which is the transport layer protocol used by vanilla Tor. In \sol, clients and servers can negotiate with each other and decide which type of channel is appropriate for a specific application.

\begin{figure}[]
 \centering
 \includegraphics[width=0.8\columnwidth]{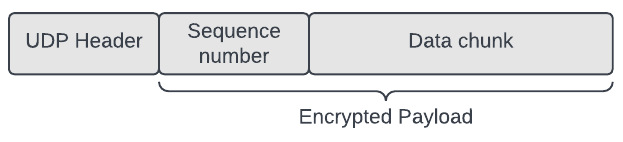}
 \caption{The format of a data packet in \sol.} 
 \label{Figure:data_packet}
 \vspace{-0.5cm}
\end{figure}

\noindent\textbf{Step 2:} After receiving a request for data, the sender of the data will encapsulate the data into multiple fixed size packets. Figure \ref{Figure:data_packet} shows the format of a UDP data packet. The payload of the packet is divided into two parts. The first part is allocated for a sequence number for each data packet and the second part contains a chunk of the actual data. The whole payload will be encrypted, and only the intended receiver of the packet will be able to decrypt it.

\noindent\textbf{Step 3:} The sender of the data will first send some metadata about the actual data (\eg number of total packets, packet sizes, allocated bytes for the sequence number, size of a data chunk) to the requested party of the data. After that, the sender will encrypt the packets and begin sending packets over the data channel. 

\noindent\textbf{Step 4:} When the receiver of the data receives a packet over the data channel, it first decrypts the packet and extracts the sequence number of the packet. Once all packets are sent over the data channel, the sender notifies the receiver over the control channel to identify the lost packet sequences.

\noindent\textbf{Step 5:} After identifying the lost packets, the receiver will send the list of lost packet sequences to the sender, and the lost packets may be retransmitted depending on the nature and needs of each application. The retransmission can done either over the data channel or over the control channel. Retransmitting lost packets over the data channel will be faster, however, it can result in subsequent packet losses if the network conditions degrade. On the other hand, retransmitting lost packets over the control channel will be slower, however, it will come with reliability due to the use of TCP between the relays of the channel.





\section {Evaluation}
\label{sec:eval}
\subsection{Evaluation Setup}

In this section, we evaluate \sol in three steps: (i) we evaluate the bootstrap time (\ie the required time to create the control channel and data channels); and (ii) we conduct experiments based on a \sol prototype that we have developed to evaluate its performance using different metrics; and (iii) we present results to understand how the performance of \sol is impacted when the number of concurrent connections increases.
We compare these evaluation results to vanilla onion services, which we use as a baseline approach. To increase the statistical significance of our results, we conducted our experiments and collected data over a period of two months.

\noindent\textbf{\sol prototype implementation:} We developed a \sol prototype in Python\footnote{We make our \sol implementation code publicly available to the research community at \url{https://github.com/malazad/Tor-with-spoofing}.}. The prototype has three main components: (i) a client module; (ii) a UDP-Tor relay module; and (iii) a server module. The client module initiates the bootstrap phase of \sol by sending an HTTP request to the onion server using the Tor Stem library \cite{tor-stem}. The server module was implemented using the Flask web development framework \cite{flask}. In both the client and the server modules, the Scrapy \cite{scrapy} framework was used to replace the actual source IP address of the sender with a temporary IP address to send packets through a data channel. Finally, the UDP-Tor relay module receives and forwards UDP packets to the next hop or to the destination. For our evaluation, we created a cloud testbed on Amazon Web Service (AWS) with instances around the world and deployed these modules on these instances.          

\noindent\textbf{Evaluation metrics:}  We have considered the following metrics for our evaluation:

\begin{enumerate} [wide, labelwidth=!, labelindent=0pt, nosep]
\item{\em Bootstrap time:} This is the time required to establish a connection between a client and a server. For vanilla onion services, it includes the time to select relay nodes by both the client and the server and the time to establish a connection through a rendezvous point. For \sol, it is the time to create the control channel and the data channels. 
\item{\em End-to-end per packet delay:} This is the time required for a packet to be delivered from a sender to a receiver. 
\item{\em Data transfer time:} This is the time required to transfer data of a certain size from a client to an onion server or vice versa. For \sol, the time to recover the lost packets is also included. 
\item{\em Packet loss:} The percentage of packets that are lost during the transmission through the data channel of \sol.
\item{\em Overhead:} The total number of bytes that need to be sent through the vanilla onion service and \sol overlay networks to successfully deliver data of a certain size. It is calculated as: $total\ number\ of\ packets \times size\ of\ each\ packet\ \times number\ of\ Tor\ relays\ between\ the\ sender\ and\ the\ receiver$. For \sol, the retransmitted packets to recover losses are also included.

\end{enumerate}


\subsection{Evaluation Results}

\subsubsection{Evaluation of the bootstrap phase of \sol}

\begin{figure}[]
 \centering
 \includegraphics[width=0.85\columnwidth]{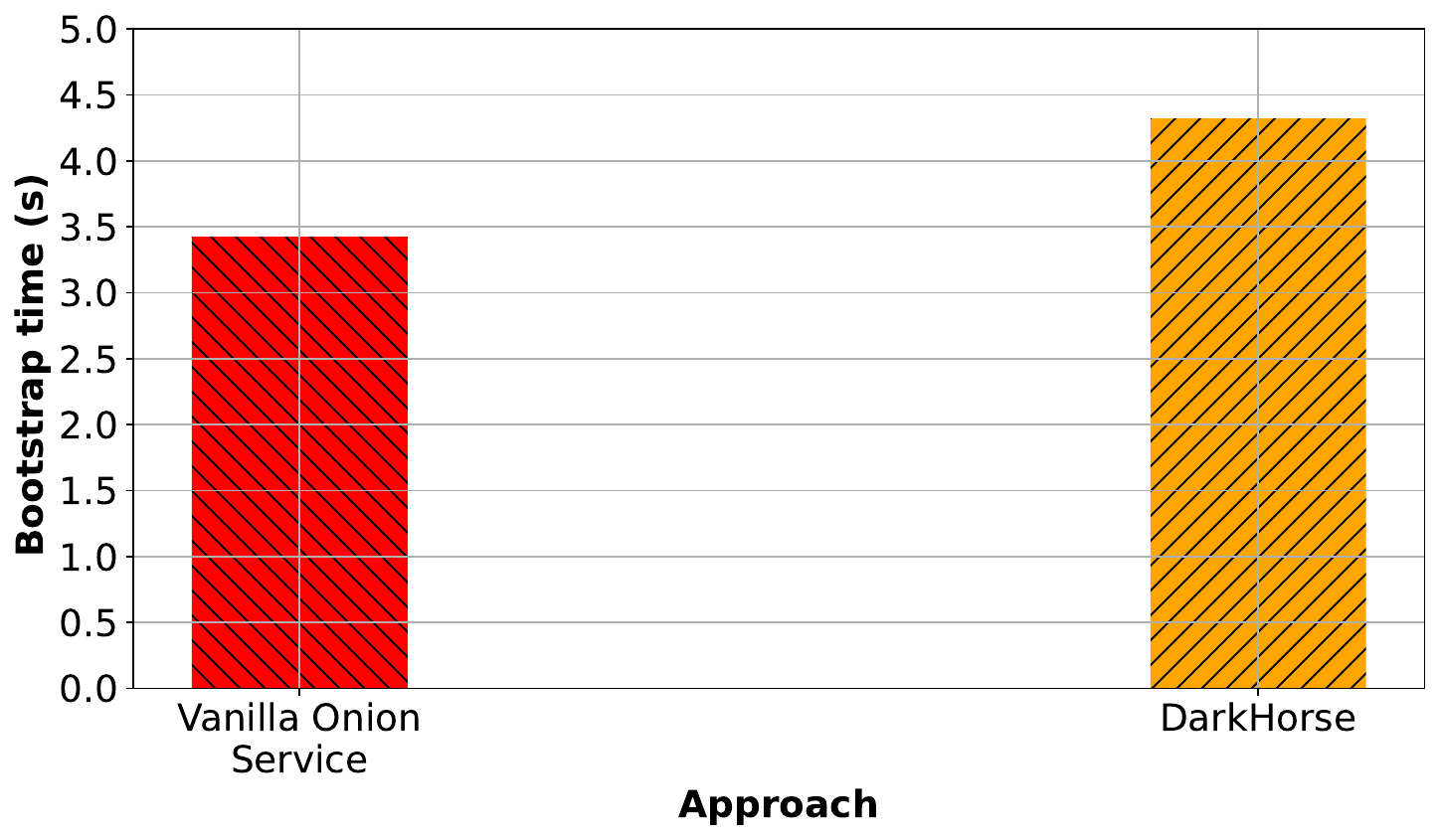}
 \caption{Comparison of bootstrap time between vanilla onion service and \sol.} 
 \label{Figure:bootstrap_time}
 \vspace{-0.5cm}
\end{figure}

\noindent\textbf{Bootstrap time:} We present the results for the bootstrap time of vanilla onion service and \sol in Figure \ref{Figure:bootstrap_time}. The evaluation results show that \sol requires around 26\% more time than vanilla onion service to establish a connection between a client and an onion server.  This extra bootstrap time overhead of \sol comes from the time required to exchange information between a sender and a receiver to create data channels. However, the additional time is overall minor (less than one second).

\begin{figure}[]
 \centering
 \includegraphics[width=0.85\columnwidth]{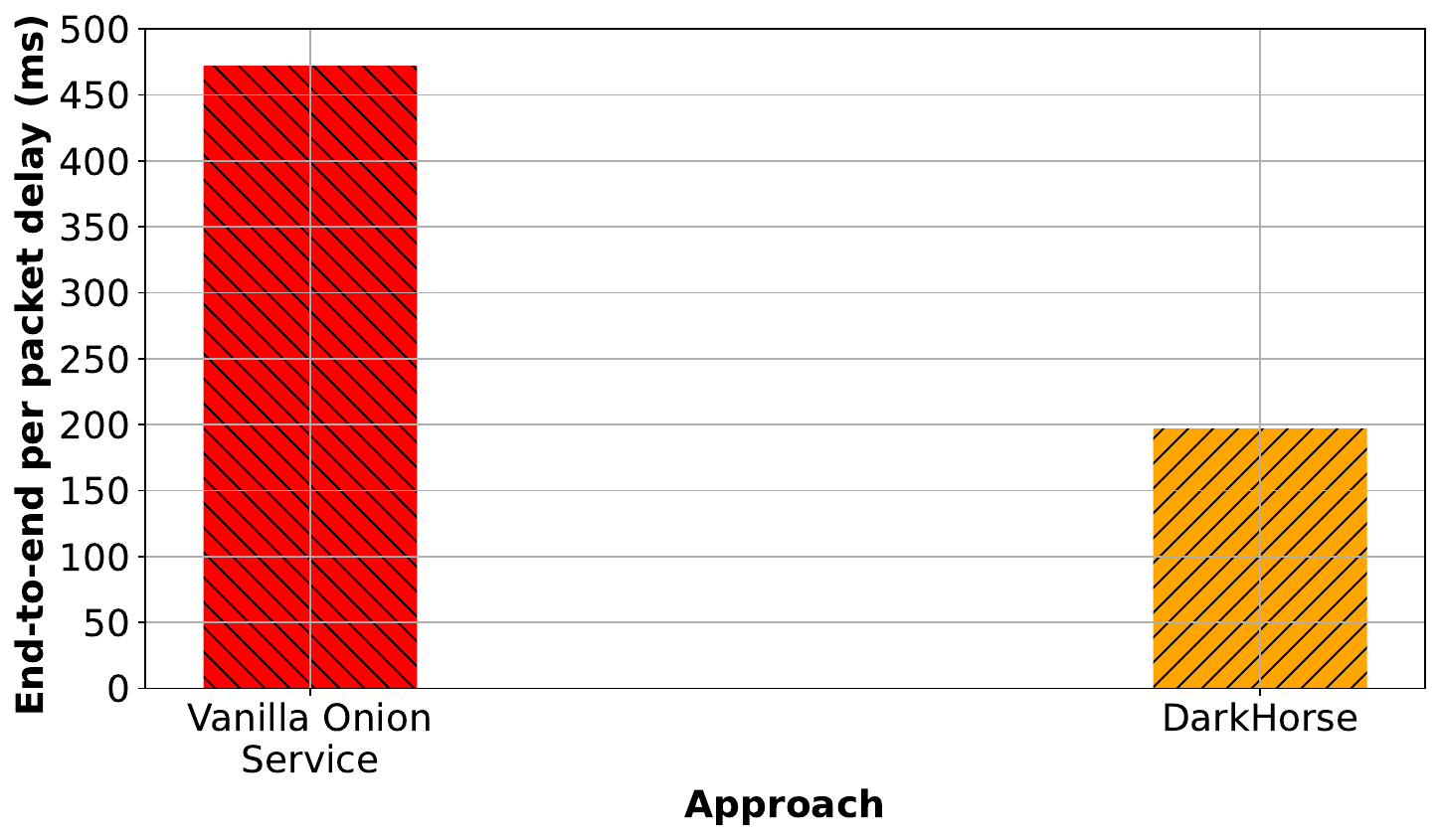}
 \caption{Comparison of end-to-end per packet delay between vanilla onion service and \sol.} 
 \label{Figure:end-to-end_time}
 \vspace{-0.5cm}
\end{figure}


\subsubsection{Evaluation of the data transfer phase of \sol}

\begin{figure}[]
 \centering
 \includegraphics[width=0.85\columnwidth]{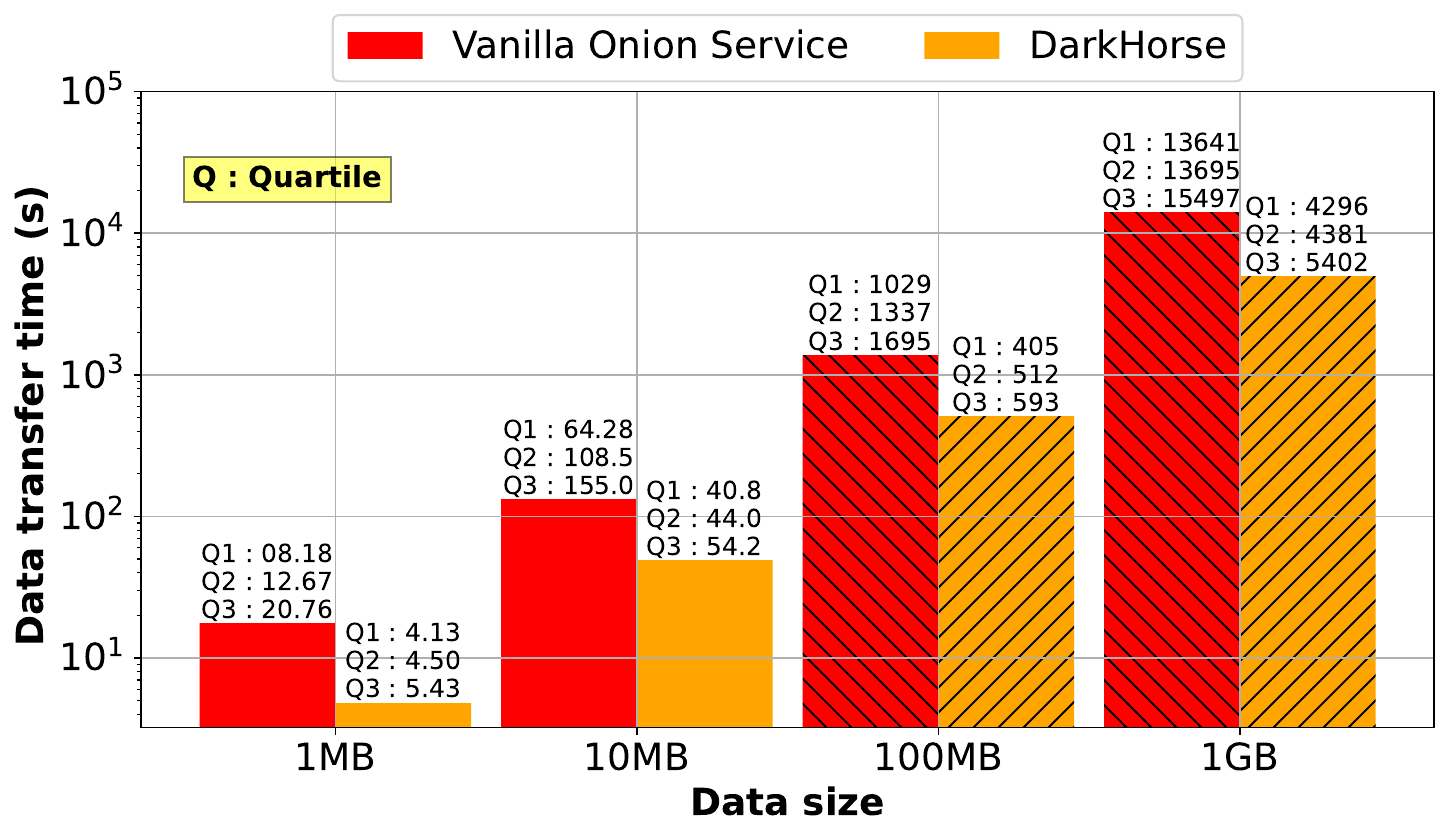}
 \caption{Data transfer time for varying data sizes.} 
 \label{Figure:time}
 \vspace{-0.5cm}
\end{figure}

\noindent\textbf{End-to-end per packet delay:} Figure \ref{Figure:end-to-end_time} shows the end-to-end per packet delay to transfer a packet from an onion server to a client using vanilla onion service and \sol. Our results indicate that \sol achieves around 58\% lower end-to-end per packet delays due to the fewer relays used along the data channel. This results in fewer encryption/decryption operations that need to be performed while delivering data from a sender to a receiver.

\noindent\textbf{Data transfer time:} Figure \ref{Figure:time} shows the average required time to transfer different sizes of data using \sol and vanilla onion service from an onion server to clients. Our results demonstrate that \sol is 2.72-3.62$\times$ faster than vanilla onion service in terms of transferring the same amount of data (including retransmissions that may be needed to recover lost packets). 
Furthermore, \sol achieves an average data transfer rate of 209.98KB/s as compared to 71.51KB/s using vanilla onion service. The quartile values also indicate that data transfer times are more consistent in \sol as compared to vanilla onion service. The speed up and the consistency in data transfer times come from the fact that \sol requires half of the Tor relay nodes compared to vanilla onion services.      

\begin{figure}[]
 \centering
 \includegraphics[width=0.85\columnwidth]{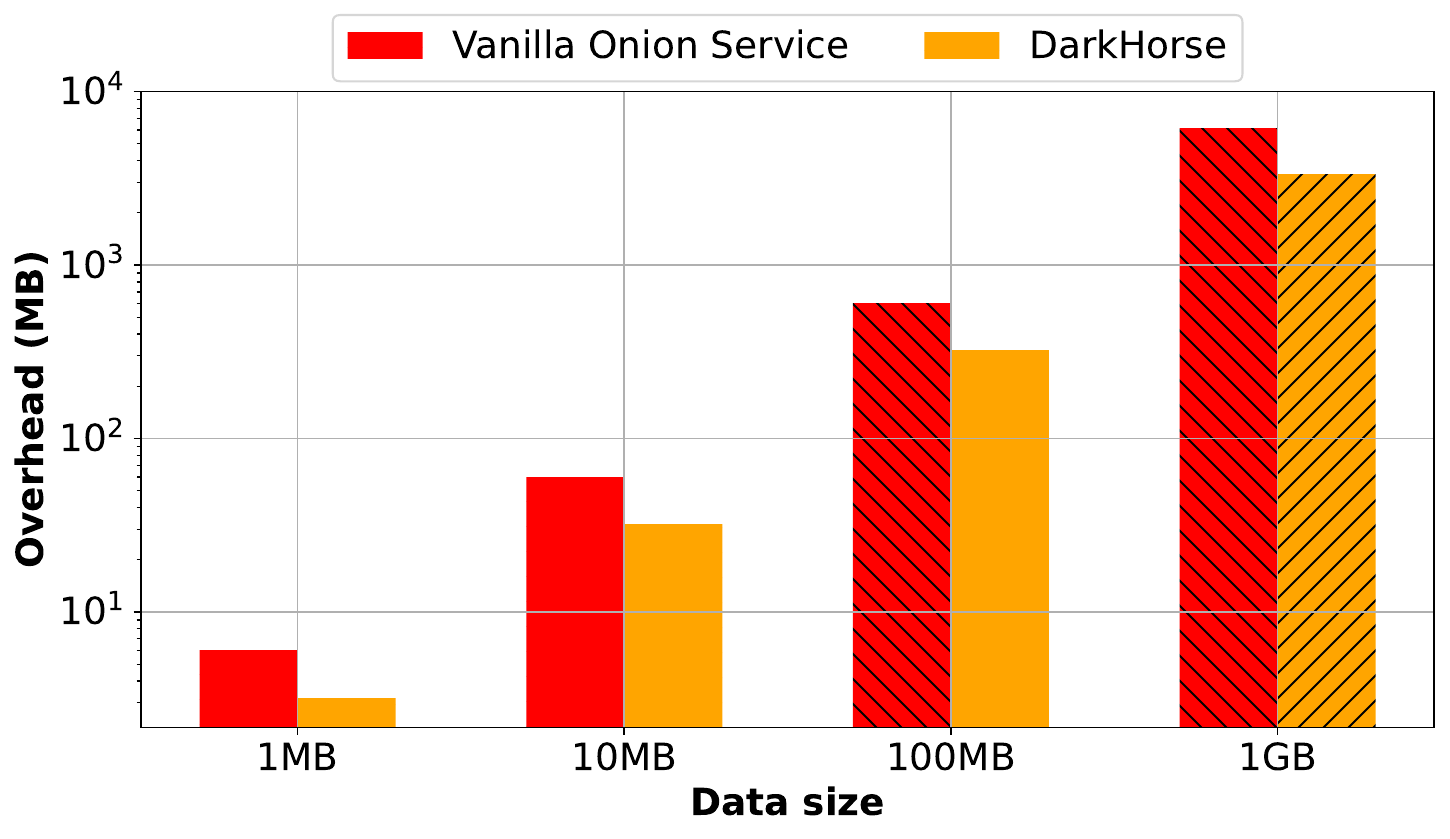}
 \caption{Overhead of vanilla onion service and \sol overlay networks for varying data sizes.} 
 \label{Figure:overhead}
 \vspace{-0.5cm}
\end{figure}

\noindent\textbf{Overhead:} In Figure \ref{Figure:overhead}, we present the results for the overhead of vanilla onion service and \sol overlay networks to transfer data of different sizes. \sol incurs about 45\%-47\% lower overhead as compared to vanilla onion service. Because of the UDP packet losses over the data channel and the retransmissions of the lost packets for recovery purposes, the overhead of \sol is not exactly 50\% less than vanilla onion service, even through the length of the data channels in \sol is half as compared to the path from a client to an onion server using vanilla onion service.

\subsubsection{Effect of concurrent data transfers on the performance of \sol} To understand how the performance of \sol changes as the number of simultaneous client data transfers grows, we increase the number of concurrent client connections up to five hundred. We discuss the results of the data transfer times and the percentage of packet losses below.

\begin{figure}[]
 \centering
 \includegraphics[width=0.85\columnwidth]{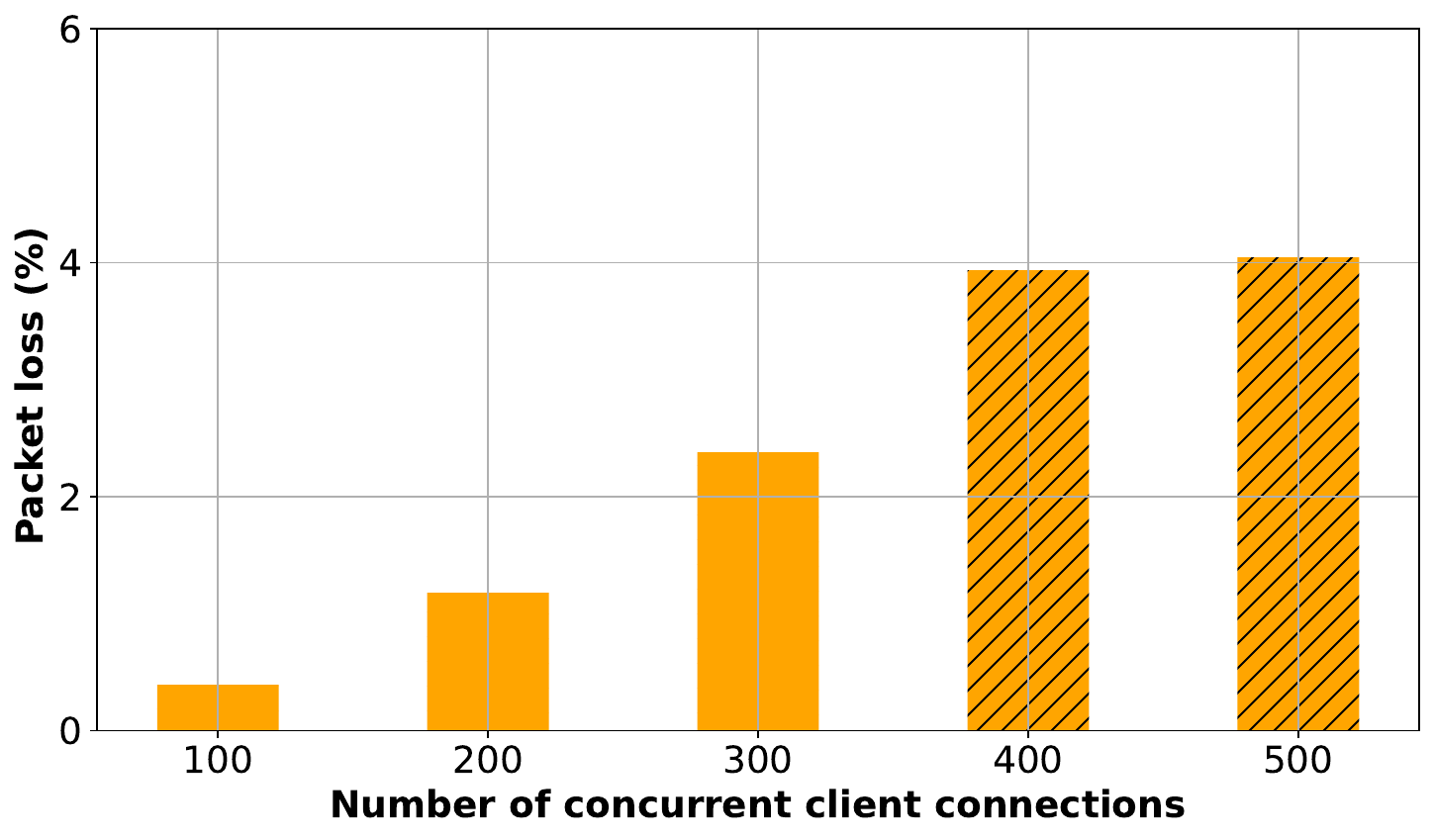}
 \vspace{-0.25cm}
 \caption{Percentage of packet loss on a data channel for varying numbers of concurrent clients.} 
 \label{Figure:dh_loss}
 \vspace{-0.55cm}
\end{figure}

\noindent\textbf{Packet losses:} In Figure \ref{Figure:dh_loss}, we present the results of packet losses on a data channel for varying numbers of concurrent client connections. Our evaluation results show that the percentage of packet losses is less than 4.04\% for up to five hundred concurrent client connections. As we increase the number of concurrent connections, the percentage of packet losses slightly increases. This increase happens because UDP does not provide congestion control mechanisms, thus senders do not dynamically adjust their data sending rates to adapt to changes of network conditions. In general, our experiments indicated that the operation of \sol can scale as much as the available resources (\eg network bandwidth, CPU and memory of relays) allow us to do.

\begin{figure}[]
 \centering
 \includegraphics[width=0.85\columnwidth]{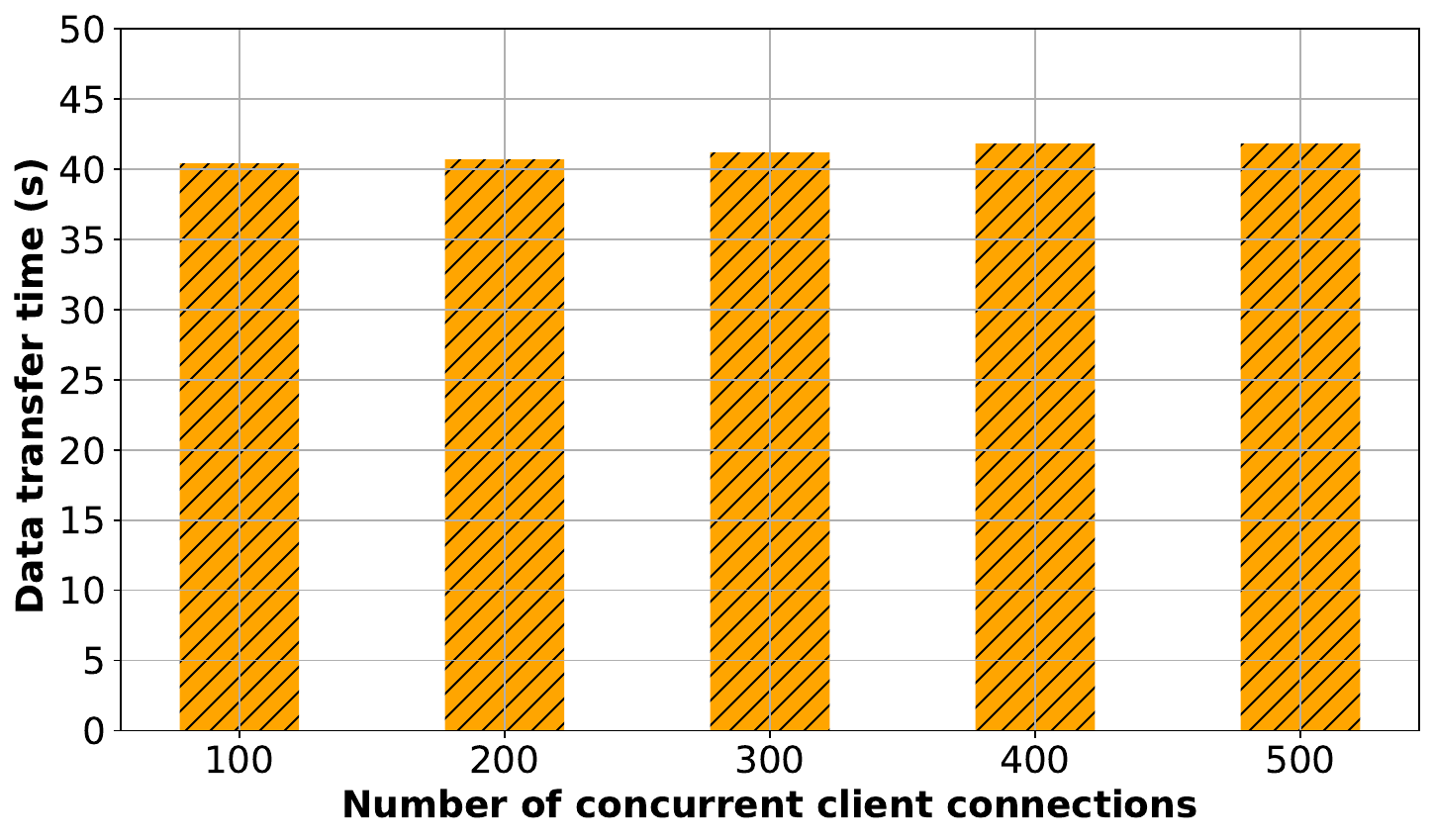}
 \vspace{-0.2cm}
 \caption{Data transfer time through a data channel for varying numbers of concurrent client connections.} 
 \label{Figure:dh_time}
 \vspace{-0.6cm}
\end{figure}

\noindent\textbf{Data transfer time:} We present the required time for transferring data for various numbers of concurrent client connections on a data channel in Figure \ref{Figure:dh_time}. As the percentage of packet losses increases with the number of concurrent client connections, the required data transfer time also increases.

\section {Discussion}
\label{sec:discussion}

\noindent\textbf{Creating a unidirectional UDP data channel:} The main goal of a unidirectional UDP data channel in \sol is to hide the identity of the sender of a packet. This can be achieved in two ways: IP spoofing \cite{ali2007ip} and Moving Target Defense (MTD) mechanisms \cite{sengupta2020survey}. In IP spoofing, the sender of a packet through the data channel will simply replace the source IP address with a different IP address (either \textit{Temp\_IP\_Server} or \textit{Temp\_IP\_Client}). In MTD mechanisms, \textit{Temp\_IP\_Server} and \textit{Temp\_IP\_Client} will be selected from a pool of IP addresses maintained by the Tor network (similar to the repository for the IP addresses of relays), therefore, a sender can request a temporary IP address from this pool. 

\noindent\textbf{Preserving anonymity in \sol during collusion:}  Let us consider the scenario of Figure \ref{Figure:data_channel}, where the onion server sends data packets to a user. In this scenario, the client knows the identity of R1, R2, and R3 but not the IP address of the server. If the user colludes with these three relays, it will not be able to identify the server since R3 receives packets from the server with a temporary IP address. On the other hand, the server only knows the IP address of R3, and R1 and R2 are unknown to it. To identify the user, the server needs to collude with R1 and R2, which are selected by the user. 

\noindent\textbf{Traffic correlation in \sol:} Unlike vanilla Tor and vanilla onion services, \sol uses asymmetric paths for sending requests and responding with data packets. As a result, it is unlikely for an attacker to observe all traffic and find traffic correlations. To find traffic correlations between a client and a server, an attacker has to monitor all three channels (two data channels and a control channel) in \sol.  

\noindent\textbf{Running experiments on a shared infrastructure:} For the evaluation of \sol, we ran experiments with virtual machine instances located around the world on a public cloud in which the resources are shared across multiple tenants. This creates an environment analogous to the vanilla Tor overlay network, where the resources of Tor relays are shared by multiple simultaneous connections.  

\section {Conclusion and Future Work}
\label{sec:conclusion}

In this paper, we presented \sol, a framework to improve the performance of onion services through the connectionless nature of UDP. \sol creates a unidirectional UDP path with a temporary sender IP address, which reduces the length of the path between a client and an onion server by half as compared to vanilla onion services. Our evaluation results demonstrated that \sol is up to 3.62$\times$ faster than vanilla onion services, while reducing the overlay network overhead by up to 47\%. In our future work, we plan to: (i) extend the \sol evaluation and make a real-world deployment of our prototype available to users; and (ii) design mechanisms to equally distribute the bandwidth of relays among concurrent connections passing through these relays. 

\section*{Acknowledgements}

This work is partially supported by the National Science Foundation (awards CNS-2104700, CNS-2306685, CNS-2016714, and CBET-2124918) and ACM SIGMOBILE.

\bibliographystyle{IEEEtran}
\bibliography{refs}

\end{document}